\documentstyle[spie]{article} 
\input{psfig}   

\def\la{\hbox{\rlap{\raise.3ex\hbox{$<$}}\lower.8ex\hbox{$\sim$}\ }}
\def\ga{\hbox{\rlap{\raise.3ex\hbox{$>$}}\lower.8ex\hbox{$\sim$}\ }}

\title{CdZnTe Background Measurement at Balloon Altitudes\\
with an Active BGO Shield} 

\author{P. Bloser\supit{a}, J. Grindlay\supit{a}, T. Narita\supit{a}, 
and F. Harrison\supit{b} 
\skiplinehalf 
\supit{a}Harvard-Smithsonian Center for Astrophysics, 60 Garden St., 
Cambridge,MA 02138, USA
\skiplinehalf 
\supit{b}Space Radiation Laboratory, 220-47 California Institute of
Technology, Pasadena, CA 91125, USA
}

\authorinfo{Further author information: (Send correspondence to
P. Bloser)\\P.B.: E-mail: pbloser@cfa.harvard.edu\\F.H.: E-mail:
fiona@srl.caltech.edu} 

\begin{document} 
  \maketitle 

\begin{abstract}
We report results of an experiment conducted in May 1997 to measure
CdZnTe background and background reduction schemes in space flight
conditions  similar to those of proposed hard X-ray astrophysics
missions.  A 1 cm$^2$ CdZnTe detector was placed adjacent to a thick BGO
anticoincidence shield and flown piggybacked onto the EXITE2
scientific balloon payload.  The planar shield was designed to veto
background counts
produced by local gamma-ray production in passive material and
neutron interactions in the 
detector.
The CdZnTe and BGO were partially surrounded by a Pb-Sn-Cu shield to
approximate the grammage of an X-ray collimator, although the field of
view was still $\sim 2\pi$ sr.  At an altitude of
127000 feet we find a reduction in background by a factor of 6 at 100
keV.  The non-vetoed background is $9 \times 10^{-4}$ cts cm$^{-2}$
s$^{-1}$ keV$^{-1}$ at 100 keV, 
about a factor of 2 higher than that of the collimated ($4.5^{\circ}$
FWHM) EXITE2 phoswich detector.  We 
compare our recorded spectrum with that
expected from simulations using GEANT and find agreement within a
factor of 2 between 30 and 300 keV.  We also compare our results
with those of
previous experiments using passive lead and active NaI shields, and
discuss possible active shielding schemes in future astronomy missions
employing large arrays of CdZnTe detectors.
\end{abstract}



\section{INTRODUCTION}
\label{sect:intro}  

The somewhat primitive state of hard X-ray and soft gamma-ray detector
technology has been given a boost in recent years by the emergence of
Cadmium Zinc Telluride (CdZnTe) as an effective wide-bandgap, high-density
semiconductor detector\cite{butler92}.  Historically, creating imaging
instruments in 
the hard 
X-ray range ($\sim$ 20--500 keV) has been especially difficult, since
energies are too high for multi-layer focusing optics (\la 80 keV) and
too low for Compton telescopes (\ga 500 keV).  The only practical
method for imaging in this band is the coded aperture technique, in
which the position of the X-ray source can be deduced from the shadow
cast by a specially-designed mask onto a large-area,
position-sensitive detector.  Although successfully employed in many
experiments, telescopes based on this principle have suffered from the
poor spatial and energy resolution of the scintillator detectors (NaI,
CsI, etc.) used to date.  CdZnTe is an attractive alternative since it
has the good energy resolution of a semiconductor, yet has far higher
stopping power than silicon and does not require cryogenic cooling
like germanium.  
Of concern is the poor mobility-lifetime ($\mu\tau$)
product for holes, which can result in poor charge collection and
degraded energy resolution.  Fortunately, it has been
found that employing a grid of pixels small relative
to the detector thickness results in an internal electric field
favorable for the collection of the electrons only\cite{barrett95}, a
phenomenon often referred to as the ``small pixel effect.''  As this is
precisely the geometry required for a position-sensitive
detector, pixellated arrays of CdZnTe for X-ray astronomy have enjoyed
intense scrutiny in 
recent years\cite{parsons94,stahle97,matteson97}.  Our work in
particular has been motivated by the need for a large-area hard X-ray
survey telescope, such as the Energetic X-ray Imaging Survey Telescope
(EXIST or EXIST-LITE) concept\cite{grindlay95,grindlay98}, designed to
be sensitive 
between 20 
and 600 keV.  Thus we have focused our study on
relatively thick (5 mm) CdZnTe detectors fashioned with various types
of pixellated arrays, and with these devices we have measured a
typical energy resolution 
of $\sim 4$\% at 60 keV\cite{bloser98,narita98}.

A major concern in any space-based astronomy mission is the level and
shape of
the detector background.  This is especially true of hard X-ray
instruments.  In coded aperture telescopes the collecting area is no
bigger than the detector, and 50\% of this area is blocked by the
mask.  Thick detectors such as those studied for EXIST have an even
more unfavorable ratio of detector volume to collecting area.  These
considerations are combined with the fact that astrophysical sources
are quite weak above $\sim 20$ keV, and that the detector is placed in
the high-radiation environment of the upper atmosphere or low Earth
orbit.  The result is that nearly all hard X-ray astronomy observations are
completely background dominated, and keeping the background to a
minimum is essential for obtaining high signal-to-noise.  

The physical processes that produce the primary background components
in an instrument are highly dependent on the specific materials used
in the detector and the surrounding structures.  Typically the
background in balloon payloads is due to the combination of diffuse
cosmic gamma-rays with gamma-ray photons and energetic particles resulting
from cosmic ray interactions in the atmosphere and in the payload itself.
Designing effective
shielding techniques is thus dependent on a thorough understanding of
the interaction processes that dominate when the detector and shields
are bombarded with gamma-ray photons, electrons, neutrons, and
protons.  Much work has been done to measure and characterize the background
components in heavily-used detector materials such as NaI and CsI
scintillators\cite{chara85,matteson77,dean91} and
germanium semiconductors\cite{gehrels85}.  A similar effort is
required for CdZnTe if its potential as a detector material for
astronomy is to be realized.  
Two measurements of CdZnTe background at
balloon altitudes are described in the literature, one each by groups
at Goddard Space Flight Center (GSFC)\cite{parsons96} and
Caltech\cite{harrison98}, 
though others exist\cite{slavis98}.  These experiments, described in
more detail in Sect.~\ref{sect:exp}, employed
simple passive and active shielding schemes and represent the first
attempts to measure and model the basic physical processes involved.
In this work we describe the first experiment designed to measure the in-flight
background in a CdZnTe detector in a configuration approximating a real
hard X-ray telescope.

  \section{EXPERIMENTAL CONFIGURATION} \label{sect:exp}

As a collaboration between the Harvard-Smithsonian Center for
Astrophysics (CfA) and Caltech we have constructed, tested, and flown
a single-element CdZnTe detector placed adjacent to a thick BGO
anticoincidence shield as a piggyback experiment on the second Energetic X-ray
Imaging Telescope Experiment (EXITE2)\cite{lum94} balloon payload.
The design of 
the instrument was motivated both by the results of previous CdZnTe
balloon flights by GSFC\cite{parsons96} and Caltech\cite{harrison98},
and by the requirements of an actual hard X-ray telescope.

The flight of the GSFC CdZnTe experiment PoRTIA, flown in 1995 as a
piggyback on the GRIS payload and containing both a thin (2 mm) and
thick (5 mm) CdZnTe detector, revealed a background
far higher than that of a Ge spectrometer flown
alongside\cite{parsons96}.  This 
background was markedly reduced, however, when the CdZnTe detector was
surrounded by NaI active shielding.  
Since a real astronomical instrument cannot completely surround the
detector with NaI, 
the current experiment places the CdZnTe detector directly in front of
a thick BGO crystal.  The
Caltech CdZnTe detector, flown on the GRIP-2 payload, was
passively shielded only and showed a high background which has been
successfully 
modeled as gamma-rays produced in the passive material by cosmic ray
interactions\cite{harrison98}.  Proposed hard X-ray telescopes using
CdZnTe, such as 
EXIST\cite{grindlay95}, usually require some passive material near the
detector in the form of a collimator; thus, such material was included
in the instrument (although for the balloon-borne EXIST-LITE
concept\cite{grindlay98} an active collimator may achieve lower
background). 

\begin{figure}
\begin{center}
\begin{tabular}{c}
\psfig{figure=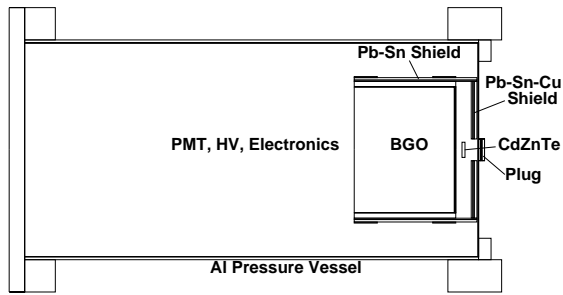,height=4cm} 
\end{tabular}
\end{center}
\caption[diagram] 
{ \label{fig:diagram}        
Schematic of the CfA-Caltech CdZnTe/BGO detector.  The CdZnTe is 1
cm$^2$ by 2 mm thick, while the BGO is 8.2 cm diameter $\times$ 6.5 cm
thick.  The detector is shielded with Pb-Sn-Cu on the front to
approximate the grammage of a passive collimator.
} 
\end{figure}

Figure~\ref{fig:diagram} shows a schematic of the CdZnTe/BGO
detector, as assembled by Caltech.  The detector is housed in an
aluminum pressure vessel with an entrance window 20 mil thick.  The
pressure vessel was filled with nitrogen at $\sim 1$ atmosphere pressure.
The CdZnTe detector was supplied
by eV Products and is a planar, single-element device 10 mm $\times$
10 mm $\times$ 2 mm.  It is 
centered about 4 mm in front of the BGO, which was supplied by JPL and
is 8.2 cm in diameter $\times$ 6.5 cm thick.  The BGO thus subtends $\sim
90$\% of the $2\pi$ steradians behind the CdZnTe.  The CdZnTe is shielded
in the front by 1.8 mm of Pb + 0.85 mm of Sn + 1.2 mm of Cu to
approximate the grammage of a graded X-ray collimator.  A hole was
left in the shield to allow calibration X-ray sources to shine in;
this hole was covered by a Pb-Sn-Cu plug attached to the outside of
the Al entrance window during flight.  The CdZnTe and BGO are shielded
on the sides by 1.6 mm of Pb + 0.9 mm of Sn.  Within the pressure
vessel, behind the BGO, are the readout PMT, high voltage supply biasing
the CdZnTe at 300 V, and an eV Products 550 preamp.  The detector is
connected to an electronics box designed and built at CfA containing
shaping amps 
(shaping time $\sim 3.3 \mu$s)
for the CdZnTe and BGO signals, discriminators, and 12-bit ADCs.  This
box also provided an interface to the main flight computer, which read
the CdZnTe/BGO data into the EXITE2 payload's data stream.  The
CdZnTe/BGO detector was mounted during flight on the side of the
EXITE2 gondola $\sim 1.5$ m above the primary phoswich detector with the
thin Al window pointed straight up.

The CdZnTe/BGO detector was extensively tested and calibrated in the
lab at CfA prior to flight using $^{241}$Am and $^{133}$Ba radioactive
sources.  We determined the energy range to be $\sim$ 15--500 keV
with an energy resolution (FWHM) of 6.2 keV at 60 keV (10.3\%).  These values
were quite stable over periods of time longer than a normal balloon
flight (\ga 24 hours).  The threshold
of the BGO shield discriminator was calibrated and set at $\sim 30$
keV with an uncertainty of 20\%.  The discriminator is not sharp, but
rolls off over $\sim 10$ keV.

  \section{FLIGHT RESULTS} \label{sect:flight}

The EXITE2 balloon payload, carrying the CdZnTe/BGO detector, was
launched from Ft. Sumner, NM at UT 16:15, May 7, 1997.  Higher than expected
winds forced the flight to be terminated after 14 hours at float
altitudes.  The altitude varied between 113000 and 127000 feet during
the night, corresponding to overlying atmospheric levels between $\sim 6$ g
cm$^{-2}$ and $\sim 3.5$ g cm$^{-2}$.  The flight computer system
behaved erratically early in the flight, but performed quite well
during the second half, and $\sim 6000$ seconds of good CdZnTe data
were collected at the highest altitudes with an average atmospheric
grammage of 3.55 g cm$^{-2}$, as well as $\sim 4100$ seconds at 5.1 g
cm$^{-2}$ and $\sim 7300$ seconds at 6 g cm$^{-2}$.  After the flight
the CdZnTe detector 
was re-tested in the lab, and the energy calibration was found to be
the same as before launch.  Thus we believe the detector performance
was quite stable during the flight.  

\begin{figure}
\begin{center}
\begin{tabular}{c}
\psfig{figure=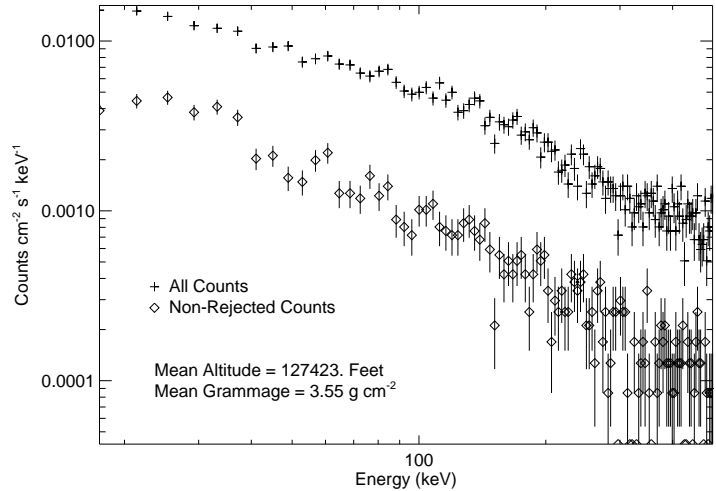,height=7cm} 
\end{tabular}
\end{center}
\caption[flight] 
{ \label{fig:flight}        
Spectrum recorded by the EXITE2 CdZnTe/BGO detector during flight at an
average atmospheric overburden of 3.55 g cm$^{-2}$.  Both total events
and non-vetoed events are shown.  The non-vetoed event rate is $9
\times 10^{-4}$ cts cm$^{-2}$ s$^{-1}$ keV$^{-1}$ at 100 keV, which is a
factor of $\sim 6$ below the total rate.
} 
\end{figure} 

Figure~\ref{fig:flight} shows the EXITE2 CdZnTe spectrum recorded from 6006
seconds of data at an average grammage of 3.55 g cm$^{-2}$,
corresponding to an average altitude of 127423 feet.  Both the total
events and the ``good,'' or non-vetoed, events are plotted.  The good
event rate at 100 keV is $\sim 9 \times 10^{-4}$ cts cm$^{-2}$
s$^{-1}$ keV$^{-1}$, roughly midway between the totally
actively-shielded GSFC (PoRTIA) 
spectrum\cite{parsons96} and the totally passively-shielded Caltech (GRIP-2)
spectrum\cite{harrison98} (see Fig.~\ref{fig:compspec}).  
The uncertainty in this background level due to data telemetry throughput
uncertainties is less than 5\%.
The good event rate is a factor of 6 
lower than the total event rate at 100 keV, showing that the planar
BGO shield placed behind the CdZnTe is an effective means of reducing
background.  The spectra recorded at 5.1 g cm$^{-2}$ and 6 g cm$^{-2}$
are essentially identical to the spectrum in Fig.~\ref{fig:flight} at
3.55 g cm$^{-2}$, ruling out significant effects due to atmospheric
depth or time-dependent activation.  The EXITE2 CdZnTe flight spectrum
will be compared in 
detail to the PoRTIA and GRIP-2 spectra in Sect.~\ref{sect:discuss}.

  \section{BACKGROUND SIMULATIONS} \label{sect:sim}

We have simulated the background recorded by the EXITE2 CdZnTe/BGO
detector at
3.55 g cm$^{-2}$ 
using the CERN Program Library simulation package GEANT.  We were
able to re-create the geometry of the 
detector quite accurately, using cylinders, tubes, and boxes of the
appropriate materials.  (Figure~\ref{fig:diagram}, in fact, was
produced by GEANT.)  Simulations were run recording the energy
deposited in both the CdZnTe and the BGO so that both the total and
non-vetoed spectra could be reproduced.  In this section we detail our
simulations and present results.

\subsection{Simulating CdZnTe Response}

In order to simulate accurately the spectrum recorded in the detector,
we had to take into account the response of CdZnTe, specifically
charge trapping.  Charge is lost from an event when the charge
carriers fall into deep traps within the crystal, resulting in low
energy tails on recorded spectral lines.  The effects of charge
trapping on the measured pulse 
height are described by the Hecht equation\cite{hecht32}, which
relates the amount of charge collected at the electrodes of a
semiconductor device to the depth at which the charge carriers are
created, the electric field in the detector, and the mobility-lifetime
products for the holes ($\mu_h\tau_h$) and 
electrons ($\mu_e\tau_e$).  Thus we record all the positions at which
energy is deposited 
in the CdZnTe detector during an event, scale each energy according to
the Hecht relation, and add up the scaled energies to give the total
effective recorded energy for that event.  The resulting spectra are
then convolved with the detector energy resolution, taken to be a
Gaussian with a width of the form FWHM $= 6.2(E/60)^{1/2}$ keV.

The electron mobility and lifetime in CdZnTe
have been measured quite accurately in the recent literature\cite{he98},
and we adopt a typical value of $\mu_e\tau_e = 5 \times 10^{-3}$
cm$^2$ V$^{-1}$.  The hole parameters are much harder to determine
since the holes are so easily trapped.
Until recently, values in the literature for $\mu_h\tau_h$ were roughly
a factor of 10 less than those for $\mu_e\tau_e$\cite{luke95}.  Using
$\mu_h\tau_h = 5 \times 10^{-4}$cm$^2$ V$^{-1}$, however,  produced
simulated spectra that 
did not show the amount of low energy tailing seen in measured
spectra.  We determined the best value of $\mu_h\tau_h$ to use by
comparing the photopeak efficiencies of the simulated spectra with
those of the measured spectra.  We fit the X-ray line spectra with a
combination of a Gaussian plus an exponential low energy tail, and
defined the photopeak efficiency as the ratio of the number of counts
under the Gaussian to the total number of
counts\cite{bloser98,narita98}.  Using 
this method we find reasonable agreement for $\mu_h\tau_h = 3 \times
10^{-5}$cm$^2$ V$^{-1}$.  
\begin{figure}
\begin{center}
\begin{tabular}{c}
\psfig{figure=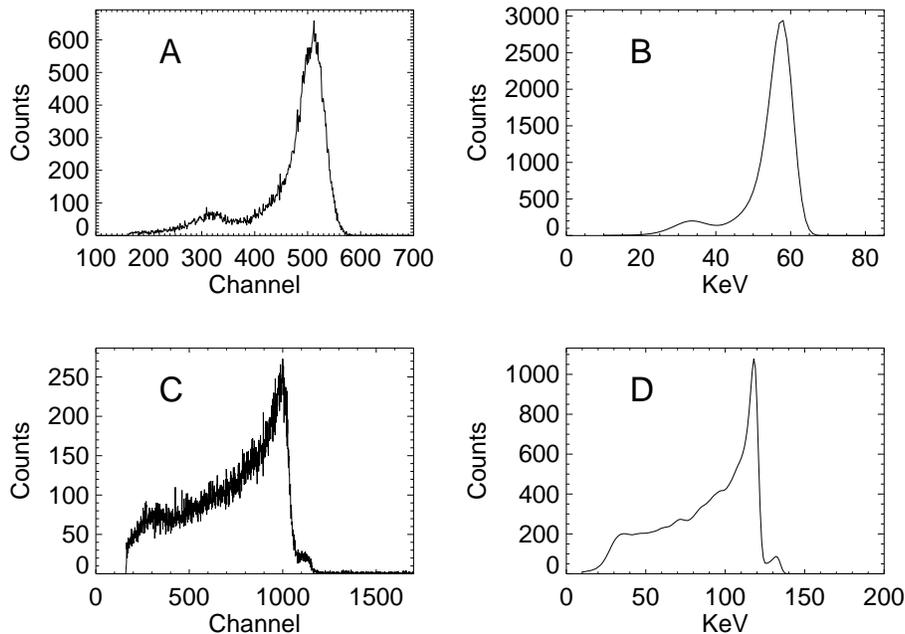,height=9cm} 
\end{tabular}
\end{center}
\caption[cztsim] 
{ \label{fig:cztsim}        
Comparison of recorded and simulated CdZnTe spectra.  A) Recorded 60
keV spectrum from $^{241}$Am.  B) Simulated $^{241}$Am spectrum.  C)
Recorded 122 keV and 136 keV $^{57}$Co spectrum.  D) Simulated
$^{57}$Co spectrum.  Here $\mu_e\tau_e = 5 \times 10^{-3}$
cm$^2$ V$^{-1}$ and $\mu_h\tau_h = 3 \times 10^{-5}$cm$^2$ V$^{-1}$.
} 
\end{figure} 
We show typical examples in Fig.~\ref{fig:cztsim}, comparing recorded
$^{241}$Am and $^{57}$Co spectra to the simulations.  For the 60 keV
line from $^{241}$Am we find photopeak efficiencies of 76\% for both the
recorded spectrum and the simulated spectrum, while for the 122 keV
line from $^{57}$Co we find recorded and simulated efficiencies of
29\% and 26\%, respectively.  
We note that such low photopeak efficiencies are the result of this
detector being a single-element device; a pixellated detector operating
in the ``small pixel'' regime (a 5 mm thick detector with 1.5 mm
pixels) gives an efficiency at 60 keV of
$\sim 90$\%, as we have reported
elsewhere\cite{bloser98,narita98}.  
Such a low value of $\mu_h\tau_h$ has
been found elsewhere when comparing simulated charge transport
to measurements of CdZnTe\cite{tous97}.

\subsection{Background Components}

In this section we describe the various components to the CdZnTe/BGO
instrument 
background that we simulate in our GEANT code.  Background is produced
by cosmic ray interactions in the atmosphere and payload material,
generating both gamma-rays and energetic particles.

\subsubsection{Shield Leakage from Atmospheric and Cosmic Diffuse Gamma-Rays}
\label{sect:leakage}

Instruments at balloon altitudes are irradiated by gamma-rays from the
diffuse cosmic X-ray and gamma-ray backgrounds and from the
atmosphere.  The atmospheric gamma-rays are produced mainly by
bremsstrahlung from electrons which are in turn the result of cosmic
ray interactions (the decay of charged mesons or the decay of neutral
pions followed by pair production).  These gamma-rays can penetrate
shielding and interact in the detector or surrounding material by
photoelectric absorption, Compton scattering, and pair production.
All of these processes can generate background counts in the detector,
and this we refer to as ``shield leakage.''

The total cosmic and atmospheric gamma-ray flux at balloon altitudes
has been measured by many 
experiments and summarized in a paper by Gehrels\cite{gehrels85}.  We
adopt the Gehrels parameterization at 3.5 g cm$^{-2}$ as the input
spectrum to our GEANT 
simulations.  This spectrum depends on zenith angle, slightly
increasing in flux for upward-going gamma-rays since much of the production
occurs in the air below the payload.  The parameterization is
uncertain by a factor of 1.4-2, especially above a few 100
keV\cite{harrison98}.  A parameterization is also given for a depth of 5
g cm$^{-2}$, but it differs from the 3.5 g cm$^{-2}$ spectrum by only
$\sim 20$\%, and then only at 10 MeV, where the uncertainties approach
a factor of 2.  We therefore do not expect the
recorded spectrum at 5 g cm$^{-2}$ to differ significantly from the
spectrum at 3.5 g cm$^{-2}$, and this agrees with our observations
(see Sect.~\ref{sect:flight}).  We restrict our modeling to the 3.5 g
cm$^{-2}$ case for simplicity, however.
The gamma-ray photons are 
introduced on the outside of the aluminum pressure vessel and allowed
to interact via the processes listed above in all the materials in the
detector and surrounding structures. The total energy deposited
in both the CdZnTe and the BGO for each event are finally recorded.

\subsubsection{Locally Produced Gamma-Rays}

The same processes that produce gamma-rays in the atmosphere also
generate photons in the payload materials, namely bremsstrahlung of
electrons from the atmosphere as well as secondary electrons produced in the
payload mass by cosmic rays.  (Electrons generated in the payload by external
gamma-rays are included in the shield leakage component, described above.)
Interactions in active
shielding materials will generally be self-vetoed, but gamma-ray
production in passive materials will contribute to the background as
long as the path of the primary electron through the passive material
does not also take it through 
an active shield.
Since the electron fluxes at balloon altitudes are poorly-determined
and it is difficult to calculate their production by cosmic ray
interactions in the
payload, 
we adopt the common practice of assuming that the gamma-ray 
production by electrons in different materials is the same as that at
a given depth in the
atmosphere, with a normalization depending on the material's atomic
number.  We use the gamma-ray production per unit mass of air given by
Dean\cite{dean89} and scale it according to the material's radiation
length in g 
cm$^{-2}$\cite{harrison98}.  The photons are then produced throughout
the passive 
materials in the detector and propagated using GEANT as before.  In
order to reproduce both the total and the non-vetoed spectra, a 
geometric correction is applied to reduce the non-vetoed
counts by the fraction of solid angle at the passive material that is
subtended by the BGO.  The main EXITE2 phoswich detector was
approximated as an 85 kg block of aluminum 1.5 m below the pressure vessel to
see whether local production in the gondola structures contributed to
the background. 

\subsubsection{Neutron-Induced Background}
\label{sect:neutronback}

Another product of cosmic ray interactions in the atmosphere and in
the payload is a flux of neutrons that may interact in the detector
due to inelastic scattering or nuclear reactions.  The background due
to elastic scattering has been found to be negligible in CdZnTe due to
the high atomic weight of Cd and Te\cite{harrison98}.  Neutron effects
become more important in experiments using thick active shielding,
since the gamma-ray contributions are effectively reduced.
Since they are neutral particles, neutrons may penetrate such
shielding without generating a veto.
The nuclear (n,$\gamma$) reaction has a
high cross section in Cd for thermal neutrons below 0.4 eV:  the
nucleus absorbs a neutron and is excited to $\sim 8$ MeV (the neutron
binding energy) above the
ground state.  The nucleus de-excites immediately, emitting $\sim 4$
gamma-rays which may then Compton scatter in the CdZnTe.  The thermal
neutron flux from the atmosphere is small (roughly $10^{-3}$ neutrons
cm$^{-2}$ s$^{-1}$ at 0.5 eV\cite{armstrong73}, which, if each
produces 4 gamma-rays at 2000 keV that Compton scatter into a flat
continuum below this energy, will generate only $\sim 2 \times
10^{-7}$ cts cm$^{-2}$ s$^{-1}$ keV$^{-1}$ in the detector), and so we
do not expect the 
(n,$\gamma$) reaction to contribute much to the detector background.
If, however, there were a local source of thermal neutrons, this
reaction might become important.  We discuss this, as well as a crude
measurement we made on this flight of atmospheric neutron fluxes, in
a forthcoming paper.
Inelastic scattering excites the nucleus 
of the target atom through the (n,n$^{\prime},\gamma$) reaction, and
$\sim 3$ de-excitation photons are emitted with mean energies lower
than those generated in the (n,$\gamma$) reaction.  Both of these
reactions emit ``prompt'' gamma-rays which may be vetoed by an active
shield.  Radioactive isotopes may
be produced by nuclear reactions (``activation'') which result in
delayed gamma-rays 
that cannot be vetoed by active shielding;
although this component has been studied previously\cite{harrison98},
it is complicated and not a major factor below several hundred keV, so
we neglect it here.  

The neutron contribution to the background is
difficult to estimate accurately, and since previous attempts to model
CdZnTe flight data with active shielding using neutron calculations
have proved difficult\cite{harrison98},
we do not attempt to calculate this component in detail here.  Rather,
we compare our 
gamma-ray calculations to the observed spectrum to see whether neutron
effects can reasonably be responsible for the difference.

\subsection{Simulation Results}

\begin{figure}
\begin{center}
\begin{tabular}{cc}
\psfig{figure=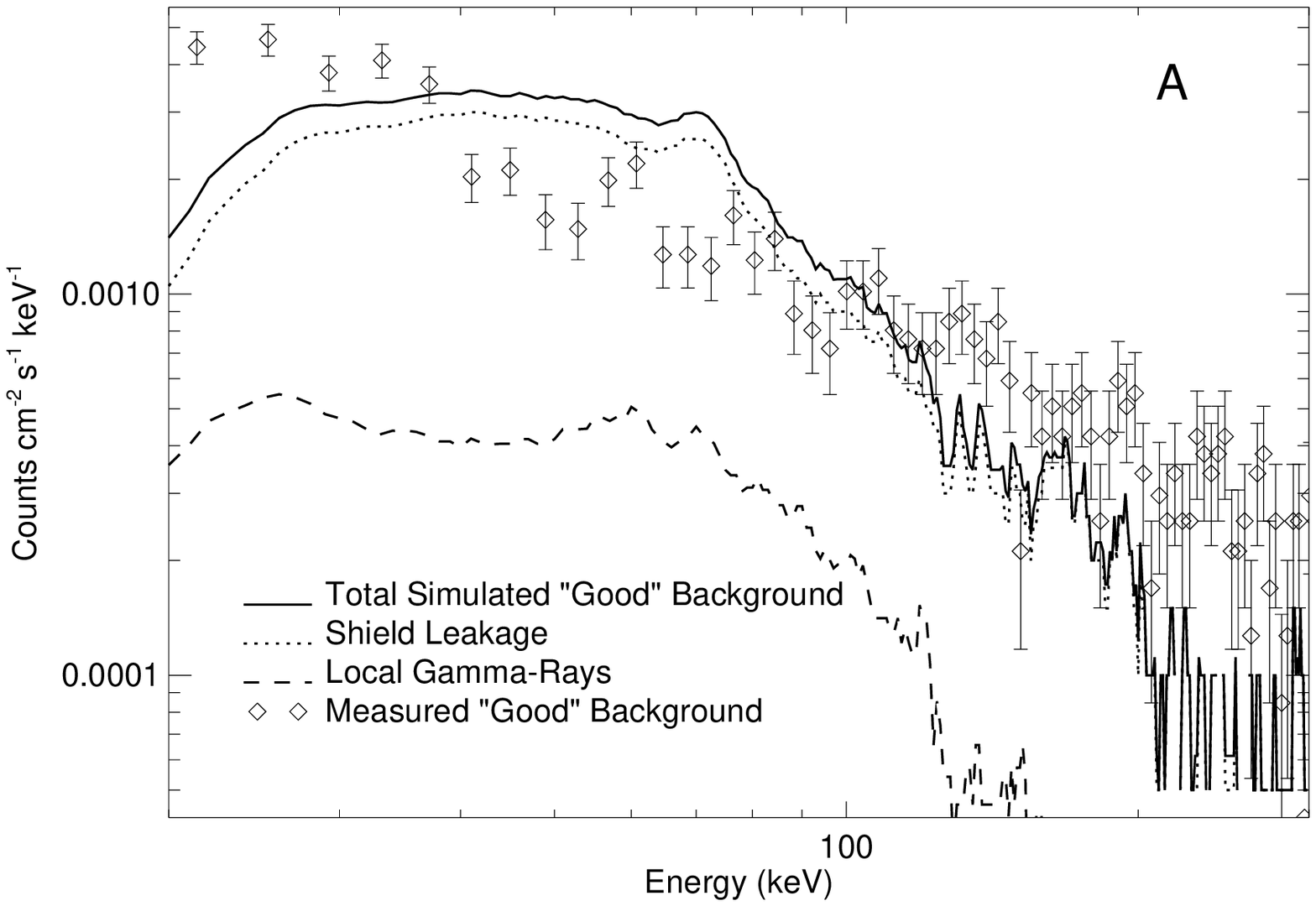,height=6cm} 
\psfig{figure=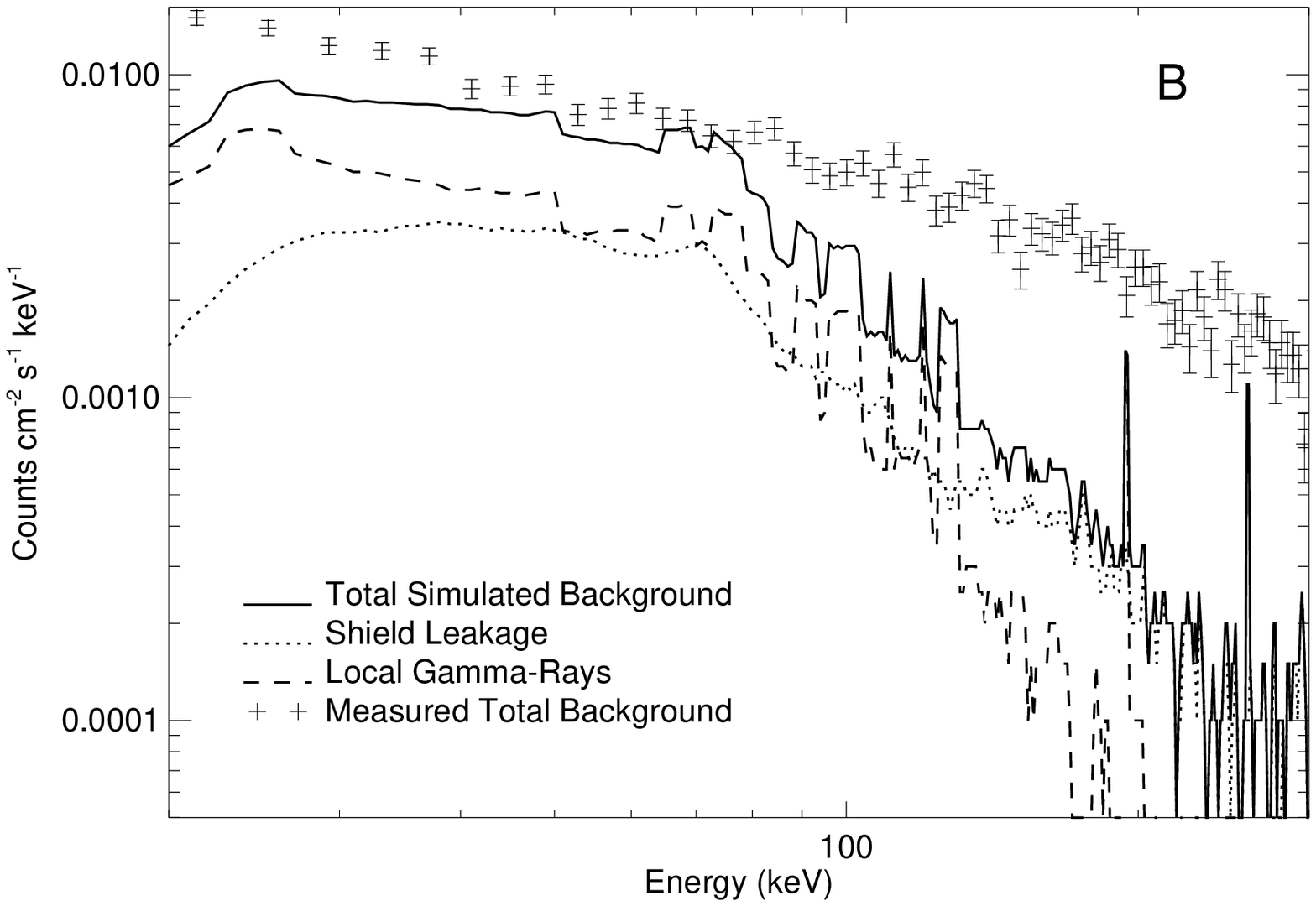,height=6cm} 
\end{tabular}
\end{center}
\caption[comp_sim] 
{ \label{fig:comp_sim}      
Comparison of the simulated spectra to the measured flight spectra for
the EXITE2 CdZnTe/BGO detector.  A) ``Good'' events (not vetoed by the
BGO).  B) All events.
The contributions from shield leakage and local gamma-rays are
indicated by the dotted and dashed lines, respectively.  
} 
\end{figure}

The results of our GEANT simulations of the gamma-ray components of
the CdZnTe/BGO detector background are shown in
Fig.~\ref{fig:comp_sim}, compared to the actual measurements, between
20 and 300 keV.
Fig.~\ref{fig:comp_sim}a shows the ``good,'' or non-vetoed spectra,
and Fig.~\ref{fig:comp_sim}b shows the total spectra.  A CdZnTe event was
considered to be vetoed if more than 30 keV was deposited anywhere in
the BGO.  The
contributions from shield leakage and local gamma-ray production are
shown separately.  The non-vetoed
background (Fig.~\ref{fig:comp_sim}a) is dominated by shield
leakage, and the predicted level agrees well (within a factor of 2)
with the data 
between 30 and 300 keV.  There is clearly an excess in the
data below 30 keV compared to the simulation.  
Above $\sim 100$ keV the simulated
spectrum seems to fall off more rapidly than the measured spectrum,
but at 300 keV they are still within a factor of 2 of each other.
There is a small peak 
in the predicted shield leakage spectrum near 70 keV, presumably due
to Pb fluorescence.  Though not discernible in the data, this
indicates that Compton and photoelectric interactions of the primary
gamma-rays in the passive material are contributing to the background.
The total background
predicted (Fig.~\ref{fig:comp_sim}b) is well within a factor of 2 of that
measured between 30 and 100 keV, and then it too falls off more rapidly than
the recorded spectrum.  
Shield leakage and local gamma-ray production 
have roughly equal magnitudes throughout the energy range shown, with
local production starting to dominate at low energies.
There is also evidence of a low energy excess 
in the data compared to the simulation.  
It appears that the total background has been
underestimated by gamma-ray effects alone, especially above 100 keV. 
The simulated total spectrum
is only a factor of $\sim 2$ greater than the simulated good
spectrum at 100 keV, as opposed to a factor of $\sim 6$ difference in
the data.  We find that local production in the main EXITE2 phoswich
detector does not produce a significant number of counts
in the CdZnTe, since it subtends a small solid angle and is below the BGO.

   \section{DISCUSSION AND CONCLUSIONS} \label{sect:discuss}

The rough agreement between the simulations and the flight CdZnTe/BGO
data is encouraging: the measured and simulated good spectra lie within a
factor of 2 of each other below 300 keV.  Considering that the input
atmospheric spectra are uncertain by factors approaching 2, especially
at higher energies, this agreement is satisfactory.  The results
indicate that local gamma-ray production in the passive material has
been effectively vetoed even by such a simple active shielding
configuration.  With more care, this component should be rendered
completely negligible in future instruments.
An active
collimator, as suggested for the EXIST-LITE concept\cite{grindlay98},
should further reduce this source of background, as well as eliminate
the contributions from fluorescence and Compton scattering.

The factor of 6 reduction in background achieved by BGO tagging is not
reproduced by the simulations.  This, together with the excesses at
high and low energies seen in the data, and the fact that the total
background is underestimated, indicates that additional components are
present in the CdZnTe background that are not accounted for by
gamma-ray interactions alone.  These components are effectively
rejected by vetoes from the BGO shield, although there is a low-energy
excess seen in the good events spectrum as well.  
It is difficult to attribute unambiguously this good-event low-energy
excess to 
additional background components, however, because the excess occurs
near the $\sim 10$ keV-wide threshold of the BGO shield:  low energy
events that 
deposit little energy in the shield are less likely to be vetoed,
resulting in an artificial excess in the ``good'' event spectrum.  No
such effects enter into the total spectrum, however.
\begin{figure}
\begin{center}
\begin{tabular}{c}
\psfig{figure=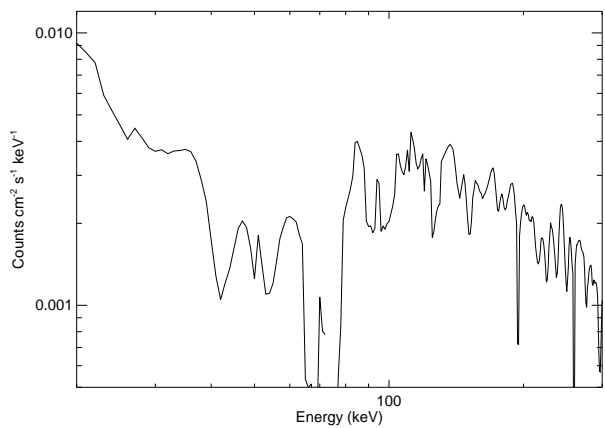,height=6cm} 
\end{tabular}
\end{center}
\caption[diff_comp] 
{ \label{fig:diff_comp}        
Residual background obtained by subtracting the total predicted
background (Fig.~\ref{fig:comp_sim}b) from the total measured
background.  There are clearly excesses at high and low energy.
} 
\end{figure} 
In Fig.~\ref{fig:diff_comp} we show the difference between the total
recorded and simulated spectra from Fig.~\ref{fig:comp_sim}b.
Although this difference is highly uncertain, there
is evidence of an excess below 40 keV, as well as between $\sim 100$ and 200
keV.  

The most obvious explanation for the difference between the
simulated gamma-induced spectrum and the data is provided by neutron
interactions.  Detailed calculations of the neutron-induced background
in CdZnTe detectors in various configurations  do
indeed indicate that prompt neutron interactions ((n,$\gamma$),
(n,n$^{\prime},\gamma$)) contribute to the background below about 50
keV\cite{harrison98}; however, this background is always at a level
far lower than that indicated by Fig.~\ref{fig:diff_comp} ($\sim
10^{-4}$ cts cm$^{-2}$ s$^{-1}$ keV$^{-1}$ at 20 keV, as opposed to
$\sim 10^{-2}$ cts 
cm$^{-2}$ s$^{-1}$ keV$^{-1}$ in the figure).  As shown in
Sect.~\ref{sect:neutronback}, the thermal component is expected to be
completely negligible unless a bright source of thermal neutrons other than
the atmosphere is present.  
Radioactive decay of the
products of neutron activation can contribute to the background up to
many hundreds of keV, but again at a level far below what is seen
here.  It is possible, however, that radioactive decay of activation
products in the BGO shield itself is producing a large high energy
background in the 
CdZnTe.  This background would certainly be efficiently vetoed.  It
remains to be seen whether more detailed estimates of the 
neutron contribution to the CdZnTe/BGO detector background can provide
an explanation for the large measured background and its marked
reduction with the BGO shield.

As noted in Sects.~\ref{sect:flight} and~\ref{sect:leakage}, we do not
observe a measurable difference between the background spectra at 3.55
g cm$^{-2}$ and 5--6 g cm$^{-2}$, nor do we expect to.  The various
observations spanned a period of 6.5 hours during the flight, and so
we can rule out any change in the background due to activation effects
over this period as well.

\begin{figure}
\begin{center}
\begin{tabular}{c}
\psfig{figure=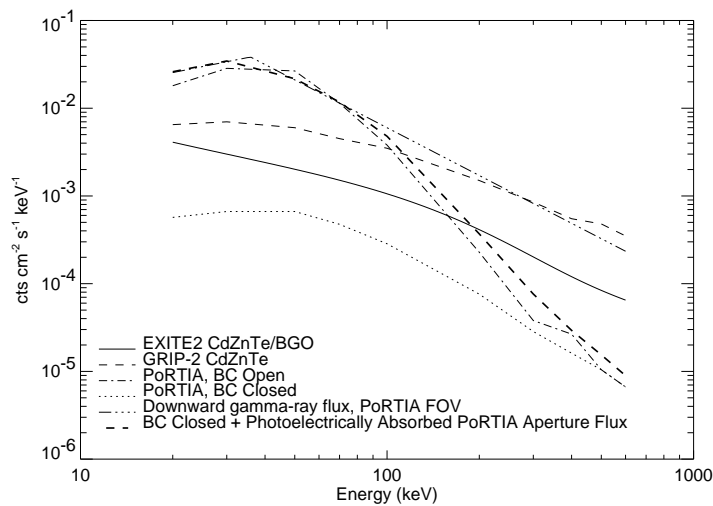,height=7cm} 
\end{tabular}
\end{center}
\caption[compspec] 
{ \label{fig:compspec}        
Comparison of CdZnTe flight spectra measured by EXITE2, PoRTIA, and
GRIP-2.  The PoRTIA and GRIP-2 spectra have been re-normalized to the
geomagnetic latitude of Ft. Sumner.  The PoRTIA spectrum is for the
thin (2 mm) detector and is shown with
the blocking crystal (BC) both 
open and closed.  The BC open PoRTIA spectrum is well-reproduced by
adding the aperture flux to the BC closed spectrum.
} 
\end{figure}

In Fig.~\ref{fig:compspec} we compare the non-vetoed CdZnTe flight background
spectra obtained by the EXITE2 CdZnTe/BGO detector, the GSFC PoRTIA
thin (2 mm)
detector\cite{parsons96}, and the Caltech GRIP-2 CdZnTe
detector\cite{harrison98}.  All three detectors were 2 mm thick, and
all were uncollimated.  The
PoRTIA and GRIP-2 balloon flights were from Alice Springs, Australia;
the data from these flights have been multiplied by 1.4 to account for
the difference in geomagnetic latitude between Alice Springs and
Ft. Sumner.  The GRIP-2 detector, as noted in Sect.~\ref{sect:exp}, was
passively shielded by Pb-Sn-Cu with thicknesses of 5 mm, 2
mm, and 2 mm, respectively (i.e., thicker than the EXITE2 CdZnTe/BGO detector
shield). 
The PoRTIA data shown were obtained with the CdZnTe
detector sitting within the GRIS anticoincidence shield so that it
was surrounded on 5 sides by 15 cm of NaI.  The top of this enclosure
could be covered with a 15 cm thick NaI blocking crystal (BC) or left
open to provide a $75^{\circ} \times 100^{\circ}$ field of view;
spectra from both modes are shown.  The background levels from the
three instruments are all within an order of magnitude of each other,
and their relative magnitudes appear to make sense: the
passively-shielded background from GRIP-2 
is the highest, the partially actively-shielded EXITE2 spectrum is
roughly between the other two, and the totally actively-shielded
PoRTIA spectrum is the 
lowest.  Their values at 100 keV are roughly $3.5 \times 10^{-3}$ cts cm$^{-2}$
s$^{-1}$ keV$^{-1}$, 
$9.0 \times 10^{-4}$ cts cm$^{-2}$
s$^{-1}$ keV$^{-1}$, and $2.8 \times 10^{-4}$ cts cm$^{-2}$
s$^{-1}$ keV$^{-1}$, respectively.  The GRIP-2 spectrum was
successfully modeled 
as being dominated by local gamma-rays produced in the passive
material\cite{harrison98}.  The simulations presented in this paper
indicate that shield leakage through the thinner lead, tin, and copper
and the subsequent gamma-ray interactions in the passive material
are the primary contributions to the background of the EXITE2 detector.
An attempt to model the BC closed PoRTIA
spectrum, including all the effects of neutron interactions,
predicted a background roughly an order of magnitude lower than that
observed below 100 keV\cite{harrison98}.  Indeed, the predicted CdZnTe
background for this case is similar to the continuum spectrum measured by the
GRIS Ge spectrometer that was flown next to PoRTIA within the NaI
shield\cite{parsons96}. 
It would seem, again, that CdZnTe suffers from internal background processes
that are not yet understood.  It is hard to see what these might be
other than neutron interactions, although the neutron calculations used
to model the PoRTIA background are quite detailed.

We are able to reproduce the PoRTIA BC open spectrum by adding to the
BC closed spectrum the photoelectrically absorbed aperture flux from
the $75^{\circ} \times 
100^{\circ}$ field of view.  This is shown in Fig.~\ref{fig:compspec}
as well.  We assume 
that any Compton scattering events in the CdZnTe will produce a veto in the
surrounding NaI.  Future balloon-borne CdZnTe telescopes with active
shielding and wide
fields of view might
use this same method to estimate their instrumental backgrounds: the
PoRTIA BC closed spectrum can be taken as an estimate of the minimum
achievable background level due to volume-dependent processes in the
absence of passive materials, and the aperture flux due to atmospheric
and diffuse cosmic gamma-rays that will be photoelectrically absorbed
at each energy can be added.  One such instrument is the EXIST-LITE
concept\cite{grindlay98}, which would use 2--4 wide field ($40^{\circ} \times
40^{\circ}$ FOV, FWHM) coded aperture telescopes with 1600--2500
cm$^2$ of CdZnTe as a detector plane to perform a sensitive all-sky
survey during an ultra-long duration balloon (ULDB) flight.  The
detector plane would be assembled from 12 mm $\times$ 12 mm pixellated
elements of CdZnTe tiled into $8 \times 8$ modules, with each module
sitting within a CsI collimator that doubles as an active
shield.  The absence of a passive collimator should reduce the
background due to Compton and photoelectric interactions, as indicated
by the EXITE2 CdZnTe/BGO results.   
\begin{figure}
\begin{center}
\begin{tabular}{c}
\psfig{figure=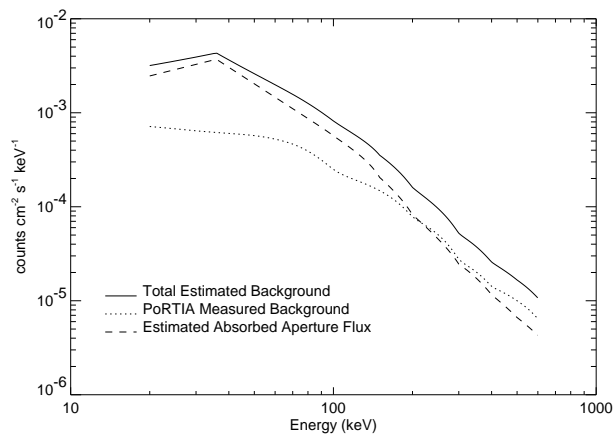,height=6cm} 
\end{tabular}
\end{center}
\caption[exist_back] 
{\label{fig:exist_back}  
Estimated background for the EXIST-LITE concept, obtained by adding
the expected aperture flux ($40^{\circ} \times 40^{\circ}$ field of
view) that is photoelectrically absorbed in 5 mm of CdZnTe to the
background measured by the thick PoRTIA detector with the blocking
crystal closed.  The 
background levels are appropriate for Alice Springs.
} 
\end{figure} 
The background estimate obtained by adding the aperture flux from a
$40^{\circ} \times 40^{\circ}$ field of view to the PoRTIA background
is shown in Fig.~\ref{fig:exist_back}.
The CdZnTe is assumed to be 5 mm thick (for response up to 600 keV),
so the PoRTIA thick (5 mm) detector spectrum was used and adjusted to
Alice Springs.  The transmission of a tantalum coded aperture mask is 
included and reduces the incident aperture flux by half at low energies.
The background
at 100 keV is $\sim 8.2 \times 10^{-4}$ cts cm$^{-2}$ s$^{-1}$ keV$^{-1}$.
This level is
roughly 30\% greater than the background measured by the EXITE2 CdZnTe/BGO
detector (the large aperture flux compensates for the lack of passive
material), and is a factor of $\sim 2.8$ higher at 100 keV than the
background 
recorded during the Ft. Sumner
balloon flight by the EXITE2 NaI/CsI phoswich detector ($\sim 4 \times
10^{-4}$ cts cm$^{-2}$ s$^{-1}$ keV$^{-1}$), rescaled to Alice
Springs.  (The EXITE2 data and other details of the flight will be
presented elsewhere\cite{chou98}.)

To make a more meaningful comparison, we consider
the background appropriate to the EXITE2 field of view
($4.5^{\circ} \times 4.5^{\circ}$ FWHM).  The absorbed aperture flux
is only $10^{-5}$ cts cm$^{-2}$ s$^{-1}$ keV$^{-1}$ at 100 keV in this
case, and so the 
instruments are not aperture dominated and the PoRTIA background may
be compared directly to the EXITE2 phoswich background.  This gives a
CdZnTe background of $\sim 3.6 \times 10^{-4}$ cts cm$^{-2}$ s$^{-1}$
keV$^{-1}$ 
at 100 keV, roughly equal to that
of the phoswich.  (These rough 
estimates obviously do not consider differences in shield leakage
between the two experiments.)
Thus, with proper shielding,  CdZnTe is a reasonably
low-background material, with levels comparable to current
scintillators.  We expect that ULDB balloon
flights or space missions employing CdZnTe detectors should be able to
achieve high-sensitivity astronomical observations in the hard X-ray
range with good spatial and spectral resolution.  If the high CdZnTe
backgrounds observed without active shielding are indeed due to
neutron interactions, one 
additional method of reducing this background even further may be the use of
``supershields,''\cite{hailey95} which use layers of low-Z material to
moderate neutrons before they are stopped by a layer of an efficient
neutron-absorber. 

We have made a measurement of the background recorded by a CdZnTe
detector flown at balloon altitudes in a configuration approximating a
plausible hard X-ray instrument.  The active BGO shield behind the
detector enables a large reduction in the CdZnTe background to a level
comparable to that of current phoswich detectors.  
Shield leakage and gamma-ray interactions in the
passive material in front of the detector make up a significant
portion of the recorded good spectrum, but
there are clearly additional components to the total CdZnTe/BGO detector
background that are effectively vetoed.  It is not clear that neutron
interactions as currently 
understood can account for the difference; it is possible that
activation in the BGO contributes at high energies.  Future hard X-ray
missions 
employing CdZnTe detectors may benefit from active collimators and
supershields.
Further effort to understand the background in CdZnTe
detectors is needed, as the good spectral and spatial resolution
possible with this material make it very valuable to the future of
hard X-ray astronomy.  We will continue our own efforts by flying a
small pixellated CdZnTe detector array surrounded by a CsI shield on the next
EXITE2 balloon flight in 1999 in preparation for
construction of a large-area CdZnTe array (EXITE3) for use in a balloon-borne
coded aperture telescope.

\acknowledgments     
 
We thank B. Matthews for work on the CdZnTe detector, T. Gauron and
J. Grenzke for work on the flight electronics and software, and K. Lum
for work on the data analysis software.  This work was supported in
part by NASA grant NAG5-5103.  P. Bloser acknowledges support from
NASA GSRP grant NGT5-50020.


  \bibliography{cztbgo_paper}   

\begin{thebibliography}{10}

\bibitem{butler92}
J.~Butler, C.~Lingren, and F.~Doty, ``Cd{Z}n{T}e gamma ray detectors,'' {\em
  IEEE Trans. Nucl. Sci.} {\bf 39}, p.~605, 1992.

\bibitem{barrett95}
H.~Barrett, J.~Eskin, and H.~Barber, ``Charge transport in arrays of
  semiconductor gamma-ray detectors,'' {\em Phys. Rev. Lett.} {\bf 75}, p.~156,
  1995.

\bibitem{parsons94}
A.~Parsons, C.~Stahle, C.~Lisse, S.~Babu, N.~Gehrels, B.~Teegarden, and P.~Shu,
  ``Room temperature semiconductor detectors for hard x-ray astrophysics,''
  {\em Proc. SPIE} {\bf 2305}, p.~121, 1994.

\bibitem{stahle97}
C.~Stahle, Z.~Shi, K.~Hu, S.~Barthelmy, S.~Snodgrass, L.~Bartlett, P.~Shu,
  S.~Lehtonen, and K.~Mach, ``Fabrication of {C}d{Z}n{T}e strip detectors for
  large area arrays,'' in {\em Hard X-ray and Gamma-Ray Detector Physics,
  Optics, and Applications},  B.~Hoover and F.~Doty, eds., {\em Proc. SPIE}
  {\bf 3115}, p.~90, 1997.

\bibitem{matteson97}
J.~Matteson, W.~Coburn, F.~Duttweiler, W.~Heindl, G.~Huszar, P.~Leblanc,
  M.~Pelling, L.~Peterson, R.~Rothschild, R.~Skelton, P.~Hink, and C.~Crabtree,
  ``{C}d{Z}n{T}e arrays for astrophysics applications,'' in {\em Hard X-ray and
  Gamma-Ray Detector Physics, Optics, and Applications},  B.~Hoover and
  F.~Doty, eds., {\em Proc. SPIE} {\bf 3115}, p.~160, 1997.

\bibitem{grindlay95}
J.~Grindlay, T.~Prince, N.~Gehrels, J.~Tueller, C.~Hailey, B.~Ramsey,
  M.~Weisskipf, P.~Ubertini, and G.~Skinner, ``{E}nergetic {X}-ray {I}maging
  {S}urvey {T}elescope ({EXIST}),'' {\em Proc. SPIE} {\bf 2518}, p.~202, 1995.

\bibitem{grindlay98}
J.~Grindlay, ``Balloon-borne hard x-ray imaging and future surveys,'' {\em Adv.
  Space Res.} {\bf 21}, p.~999, 1998.

\bibitem{bloser98}
P.~Bloser, T.~Narita, J.~Grindlay, and K.~Shah, ``Prototype imaging
  {C}d-{Z}n-{T}e array detector,'' in {\em Semiconductors for Room-Temperature
  Radiation Detector Applications II},  R.~James, T.~Schlesinger, P.~Siffert,
  M.~Cuzin, M.~Squillante, and W.~Dusi, eds., {\em Proc. MRS} {\bf 487},
  p.~153, 1998.

\bibitem{narita98}
T.~Narita, P.~Bloser, J.~Grindlay, R.~Sudharsanan, C.~Reiche, and C.~Stenstrom,
  ``Development of prototype pixellated {PIN} {C}d{Z}n{T}e detectors,'' in {\em
  Hard x-ray and gamma-ray detector physics and applications},  {\em Proc.
  SPIE} {\bf 3446}, p.~218, 1998.

\bibitem{chara85}
P.~Charalambous, A.~Dean, R.~Lewis, and N.~Dipper, ``The background noise in
  space borne low energy gamma-ray telescopes,'' {\em Nuc. Inst. Meth. A} {\bf
  238}, p.~533, 1985.

\bibitem{matteson77}
J.~Matteson, P.~Nolan, W.~Paciesas, and R.~Pelling, ``Design and performance of
  an actively collimated phoswich system for x-ray astronomy,'' {\em Space Sci.
  Instr.} {\bf 3}, p.~491, 1977.

\bibitem{dean91}
A.~Dean, F.~Lei, and P.~Knight, ``Background in space-borne low-energy
  gamma-ray telescopes,'' {\em Space Sci. Rev.} {\bf 57}, p.~109, 1991.

\bibitem{gehrels85}
N.~Gehrels, ``Instrumental background in balloon-borne gamma-ray spectrometers
  and techniques for its reduction,'' {\em Nuc. Inst. Meth. A} {\bf 239},
  p.~324, 1985.

\bibitem{parsons96}
A.~Parsons, S.~Barthelmy, L.~Bartlett, F.~Birsa, N.~Gehrels, J.~Naya, J.~Odom,
  S.~Singh, C.~Stahle, J.~Tueller, and B.~Teegarden, ``{C}d{Z}n{T}e background
  measurements at balloon altitudes,'' {\em Proc. SPIE} {\bf 2806}, p.~432,
  1996.

\bibitem{harrison98}
F.~Harrison, C.~Hailey, J.~Hong, A.~Wong, and W.~Cook, ``Background in
  balloon-borne hard x-ray/soft gamma-ray cadmium zinc telluride detectors,''
  {\em Nuc. Inst. Meth. A}, in press.

\bibitem{slavis98}
K.~Slavis, W.~Binns, P.~Dowkontt, J.~Epstein, P.~Hink, J.~Matteson,
  F.~Duttweiler, G.~Huszar, P.~Leblanc, M.~Pelling, R.~Skelton, and E.~Stephan,
  ``High altitude balloon flight of {CZT} detectors for high energy x-ray
  astronomy,'' {\em BAPS} {\bf 43}, p.~1087, 1998.

\bibitem{lum94}
K.~Lum, R.~Manandhar, S.~Eikenberry, M.~Krockenberger, and J.~Grindlay,
  ``Initial performance of the {EXITE2} imaging phoswich detector/telescope for
  hard x-ray astronomy,'' {\em IEEE Trans. Nucl. Sci.} {\bf NS-41}, p.~1354,
  1994.

\bibitem{hecht32}
K.~Hecht, ``{Z}um {M}echanismus des {L}ichtelektrischen {P}rimarstromes in
  {I}solierenden {K}ristallen,'' {\em Zeits. Phys} {\bf 77}, pp.~235--243,
  1932.

\bibitem{he98}
Z.~He, G.~Knoll, and D.~Wehe, ``Direct measurement of electron drift parameters
  of wide band gap semiconductors,'' {\em Nuc. Inst. Meth. A}, in
  press.

\bibitem{luke95}
P.~Luke, ``Unipolar charge sensing with coplanar electrodes - applications to
  semiconductor detectors,'' {\em IEEE Trans. Nucl. Sci.} {\bf 42}, p.~207,
  1995.

\bibitem{tous97}
O.~Tousignant, L.~Hamel, J.~Courville, P.~Paki, J.~Macri, K.~Larson, M.~Mayer,
  M.~McConnell, and J.~Ryan, ``Progress in the study of {C}d{Z}n{T}e strip
  detectors,'' in {\em Hard X-ray and Gamma-Ray Detector Physics, Optics, and
  Applications},  B.~Hoover and F.~Doty, eds., {\em Proc. SPIE} {\bf 3115},
  p.~214, 1997.

\bibitem{dean89}
A.~Dean, F.~Lei, K.~Byard, A.~Goldwurm, C.~Hall, and J.~Harding, ``The
  gamma-ray emissivity of the {E}arth's atmosphere,'' {\em Astron. Astrophys.}
  {\bf 219}, p.~358, 1989.

\bibitem{armstrong73}
T.~Armstrong, K.~Chandler, and J.~Barish, ``Calculations of neutron flux
  spectra induced in the {E}arth's atmosphere by galactic cosmic rays,'' {\em
  J. Geophys. Res.} {\bf 78}, p.~2715, 1973.

\bibitem{chou98}
Y.~Chou, P.~Bloser, J.~Grenzke, J.~Grindlay, K.~Lum, G.~Monnelly, and
  B.~Robbason, ``{EXITE2} detector and telescope development and initial
  balloon flight results,'' {\em IEEE Trans. Nucl. Sci.}, in
  preparation.

\bibitem{hailey95}
C.~Hailey and F.~Harrison, ``A new concept for background rejection in
  gamma-ray astronomy - the supershield,'' {\em Nuc. Inst. Meth. A} {\bf 365},
  p.~518, 1995.

\end{thebibliography}
  \bibliographystyle{spiebib}   
 
  \end{document}